\documentstyle[aps,prl,twocolumn,epsf]{revtex}
\begin{document}
%\draft
\title{Chaos and Interactions in Quantum Dots}

\author{ Y. Alhassid}

\address{Center for Theoretical Physics, Sloane Physics Laboratory,
     Yale University, New Haven, Connecticut 06520, USA}
\date {\today}
\maketitle
\begin{abstract}
Quantum dots are small conducting devices containing up to several thousand electrons.  We focus here on closed dots whose  single-electron dynamics are mostly chaotic. The mesoscopic fluctuations of the conduction properties of such dots reveal the effects of one-body chaos, quantum coherence and electron-electron interactions.\footnote{To apppear in the Proceedings of the Nobel Symposium on Quantum Chaos 2000, B\"{a}ckaskog Castle, Sweden.}
\end{abstract}

%\pacs{PACS numbers: 73.40.Gk, 05.45+b, 73.20.Dx, 73.23. -b}
\narrowtext

  \section{Quantum Dots}

\begin{figure}[h!]
\epsfxsize= 5.5 cm
\centerline{\epsfbox{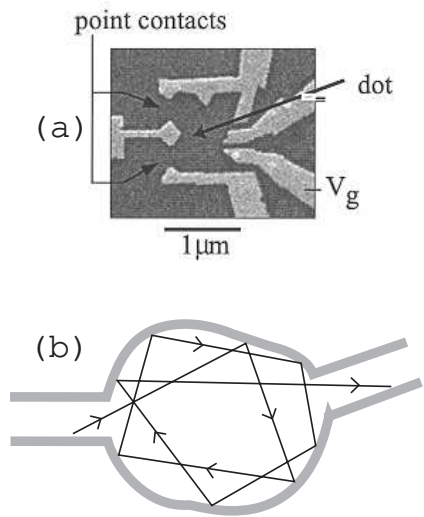}}
\vspace{3 mm}
\caption {Quantum dots: (a) a scanning electron micrograph of a dot used by Folk {\em et al.} \protect\cite{Folk96}.  A 2D electron gas is formed in the interface of a GaAs/AlGaAs heterostructure (darker area). Electrostatic potentials applied to metallic gates (lighter shade) confine the electrons to a sub-micron region. The shape and area of the dot can be changed by controlling the gate voltages  $V_{g1}$ and $V_{g2}$; (b) a schematic drawing of a ballistic dot attached to two leads. The electron's trajectory scatters from the dot's boundaries several times before exiting.
}
\label{fig1}
\end{figure}

Recent advances in materials science have made possible the fabrication of quantum dots,  submicron-scale conducting devices containing up to several thousand electrons \cite{Kouwenhoven97}.  A 2D electron gas is created at the interface region of a semiconductor heterostructure (e.g., GaAs/AlGaAs) and the electrons are further confined to a small region by applying a voltage to metal gates,  depleting the electrons under them. Insofar as the motion of the electrons is restricted in all three dimensions, a quantum dot may be considered a zero-dimensional system.  The transport properties of the dot, i.e., its conductance, can be measured by connecting it to external leads.  A micrograph of a quantum dot is shown in Fig. \ref{fig1}(a).

  At low temperatures, the electron preserves its phase over  distances that are longer than the system's size, i.e., $L_\phi > L$, where $L_\phi$ is the coherence length and $L$ is the linear size of the system.   Such systems are called mesoscopic. Elastic scatterings of the electron from impurities generally preserve phase coherence, while inelastic scatterings, e.g., from other electrons or phonons, result in phase breaking

 When the mean free path $\ell$ is much smaller than $L$, transport across the dot is dominated by diffusion, and the system is called diffusive.  In the late 1980s it became possible to fabricate devices with little disorder where $\ell > L$.  In these so-called ballistic dots, transport is dominated by scattering of the electrons from the boundaries.  A schematic illustration of a ballistic dot is shown in Fig. \ref{fig1}(b).

  In small dots (with typically less than $\sim 20$ electrons), the confining potential is often harmonic-like, leading to regular dynamics of the electron and shell structure that can be observed in the addition spectrum (i.e., the energy required to add an electron to the dot).  Maxima in the addition spectrum are seen for numbers of electrons  that correspond to filled (${\cal N} = 2, 6, 12$) or half-filled (${\cal N} = 4, 9, 16$) valence harmonic-oscillator shells \cite{Tarucha96}. 

  Dots with a large number of electrons (${\cal N} \geq 50 - 100$) often have  no particular symmetry, and their irregular shape results in single-particle dynamics that are mostly chaotic.  For such dots the conductance and addition spectrum display ``random'' fluctuations when the shape of the dot or a magnetic field are varied. This is the statistical regime, where we are interested in the statistical properties of the dot's spectrum and conductance when sampled from different shapes and magnetic fields. For a recent review of the statistical theory of quantum dots see Ref. \cite{Alhassid-rev}.

\begin{figure}[h!]
\epsfxsize= 6.5 cm
\centerline{\epsfbox{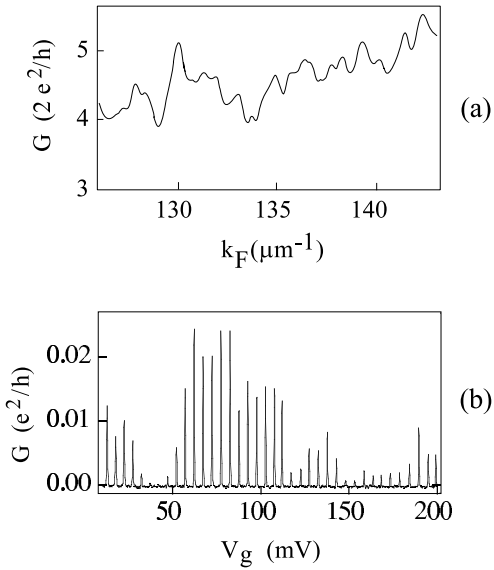}}
\vspace{3 mm}
\caption{(a) Conductance vs the electron's Fermi momentum $k_F$ in an open dot (from Ref. \protect\cite{Keller94}). (b) Conductance vs gate voltage in a closed dot displaying a series of Coulomb-blockade peaks (from Ref. \protect\cite{Folk96}). 
}
\label{fig2}
\end{figure}

 Many of the physical parameters of a quantum dot can be experimentally controlled, including its degree of coupling to the leads, shape, size, and number of electrons. When the dot is  ``open'', i.e., strongly coupled to leads, there are generally several channels in each lead and the conductance fluctuates as a function of, e.g., the Fermi momentum of the electron in the leads (see Fig. \ref{fig2}(a)). As the point contacts are pinched off, the coupling becomes weaker and a barrier is effectively formed between the dot and the leads. In such ``closed'' dots, the charge is quantized. At low temperatures, the conductance through a closed dot displays peaks as a function of gate voltage (or Fermi energy); see, e.g., Fig. \ref{fig2}(b). Each peak represents the addition of one more electron into the dot.   In between the peaks, the tunneling of an electron into the dot is blocked by the Coulomb repulsion of electrons already in the dot, an effect known as Coulomb blockade. In this paper we discuss closed dots in which mesoscopic phenomena are determined by the interplay between single-particle chaos, quantum coherence and electron-electron interactions. 

\section{ Transport in the Coulomb blockade regime}

  The simplest model for describing the Coulomb blockade regime is the constant interaction (CI) model, in which the Coulomb energy is taken to be $e^2 {\cal N}^2/2C$, where $C$ is the total capacitance of the dot and ${\cal N}$ is the number of electrons.  The Hamiltonian of the CI model is given by  
\begin{equation}\label{CI}
H_{\rm dot} = \sum_\lambda (E_\lambda-e \alpha V_g) a_\lambda^\dagger
a_\lambda^{}  + e^2 \hat{\cal N}^2/2C \;.
\end{equation}
Here $a_\lambda^\dagger | 0 \rangle$ is a complete set of single-particle
eigenstates  in the dot with energies $E_\lambda$, and $\hat{\cal N}=\sum\limits_\lambda
a^\dagger_\lambda  a_\lambda$ is the electron number operator in the dot.
 The quantity $\alpha V_g$ is the confining potential written in terms of a gate voltage $V_g$ and $\alpha
=C_g/C$, where $C_g$ is the gate-dot capacitance.

\begin{figure}[h!]
\epsfxsize= 8. cm
\centerline{\epsfbox{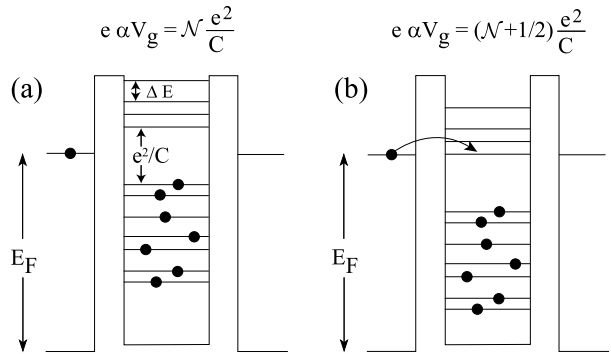}}
\vspace{3 mm}
\caption{Schematic diagram of Coulomb blockade in closed dots: (a) when the gate voltage is  $e\alpha V_g={\cal N}e^2/C$
there is a charging energy gap on both sides of the Fermi energy,  blocking the tunneling of electrons into the dot; 
 (b) when the gate voltage increases to $e\alpha  V_g=({\cal N}+ 1/2)e^2/C$, the charging energy for adding an electron to the dot vanishes.  When the Fermi energy of the electron in the lead matches the first unoccupied single-particle state in the dot (i.e., $E_F=E_{{\cal N}+1}$), resonant tunneling of an electron into the dot occurs. From Ref. \protect\cite{Alhassid-rev}.
}
\label{fig3}
\end{figure}

  At low temperatures, conductance occurs by resonant tunneling through a single-particle level in the dot. Assuming energy conservation for the tunneling of the ${\cal N}$-th electron we have, $E_F + {\cal E}_{g.s.} ({\cal N}-1) ={\cal E}_{g.s.}({\cal N})$, where $E_F$ is the Fermi energy of the electron in the leads and ${\cal E}_{gs}({\cal N})$ is the ground state energy of a dot with ${\cal N}$ electrons. Using Eq. (\ref{CI}), we find  that the effective Fermi energy
$\tilde E_F \equiv E_F + e \alpha V_g$ satisfies 
\begin{equation}\label{peak-energies}
\tilde E_F  = E_{\cal N}
 +\left( {\cal N}-\frac{1}{2}\right){e^2 \over  C} \;.
\end{equation}
The conductance displays a series of peaks at values of $\tilde E_F$ given by (\ref{peak-energies}), with each peak corresponding to the tunneling of an additional electron into the dot.  The spacings between the peaks are given by   
\begin{equation}\label{peak-spacing}
\Delta_2 \equiv \Delta \tilde E_F =  (E_{{\cal N} + 1} - E_{\cal N}) + e^2/C \;.
\end{equation}
Since the charging energy is usually much larger than the mean-level spacing $\Delta$, the Coulomb-blockade peaks are almost equidistant. Coulomb blockade is illustrated in Fig. \ref{fig3}.

   The Coulomb-blockade peak heights contain information about the wave functions. For closed dots, a typical level width $\Gamma$ is small, and $\Gamma \ll T$ even for the lowest electron temperatures attained in the experiments. Under this condition, the coherence between the electrons in the leads and the dot can be ignored, and a master-equation approach is feasible \cite{Beenakker91}.  In the CI model and for $\Gamma \ll T \ll \Delta$, the ${\cal N}$-th conductance peak  $G$  occurs through level $\lambda = {\cal N}$ and is given by \begin{equation}\label{conductance-peak}
G (\tilde E_F,T) \approx
G_\lambda^{\rm peak} {1 \over \cosh^2\left({E_\lambda - \tilde E_F \over 2kT}\right)}
\;.
\end{equation}
Eq. (\ref{conductance-peak}) describes a peak centered at $\tilde E_F = E_\lambda$ (here $\tilde E_F$ is measured with respect to $({\cal N}-1/2)(e^2/C)$). The peak has a width $\sim T$ and a height of
\begin{equation}\label{peak-height}
G^{\rm peak}_\lambda = {e^2 \over h} {\pi \bar \Gamma \over 4 kT} g_\lambda \;,\;\;\;\; {\rm where} \;\;\;\;\; g_\lambda = {2 \over \bar\Gamma} {\Gamma^l_\lambda
\Gamma^r_\lambda  \over \Gamma^l_\lambda + \Gamma^r_\lambda } \;.
\end{equation}
The quantities $\Gamma^{l(r)}$ describe the partial width of level $\lambda$ to decay into the left (right) lead, and $\bar \Gamma$ is the average level width.  

  The partial widths can be written as $\Gamma_\lambda = |\gamma_\lambda|^2$, where $\gamma_\lambda$ is the partial width amplitude.  In $R$-matrix theory \cite{Wigner47,Lane58,Jalabert92}, $\gamma_\lambda$  can be related to the wave function $\Psi_\lambda$  at the respective dot-lead interface 
\begin{eqnarray}\label{partial-amplitude}
\gamma_{c\lambda}= \sqrt{\hbar^2 k_c P_c \over m}   \int\limits_{\cal C} dl\;
   \phi_c^{\ast}\Psi_\lambda \;,
\end{eqnarray}
where $k_c$ is the longitudinal channel momentum
($\hbar^2k_c^2/2m + \hbar^2\kappa_c^2/2m = E$),
$P_c$  is the penetration factor to tunnel through
the barrier in channel $c$ ($P_c=1$ in the absence of barrier and $P_c \ll 1$ in the presence of a barrier), and $\phi_c$ is the transverse channel wave function. The integral in (\ref{partial-amplitude}) is over the dot-lead interface ${\cal C}$.

\section{Signatures of chaos in closed dots}

  In a dot where the classical dynamics of the electron are chaotic, the wave function fluctuations are described by random-matrix theory (RMT) \cite{Mehta91}. These fluctuations lead to fluctuations in the conductance peaks according to Eqs. (\ref{peak-height}) and (\ref{partial-amplitude}). When time-reversal symmetry is conserved (i.e., there is no external magnetic field $B$) the corresponding random-matrix ensemble is the Gaussian orthogonal ensemble (GOE), while for broken time-reversal symmetry ($B \neq 0$) the appropriate ensemble is the Gaussian unitary ensemble (GUE).  For a recent review of RMT and its applications see Ref. \cite{Guhr98}.  For a review of the random-matrix theory of quantum transport (including applications to open dots) see Ref. \cite{Beenakker97}.

  To quantify the conductance peak statistics using RMT, we  express the partial width amplitude (\ref{partial-amplitude}) as a scalar product of the resonance wavefunction
$\bbox{\psi}_\lambda= (\psi_{\lambda 1},\psi_{\lambda
2},\ldots)$ and the channel wavefunction $\bbox{\phi}_c =
(\phi_{c  1}, \phi_{c 2}, \dots)$
\begin{equation}\label{scalar-product}
\gamma_{c \lambda} = \langle \bbox{\phi}_c |
 \bbox{\psi}_\lambda \rangle \equiv \sum\limits_j \phi^\ast_{c j}
\psi_{\lambda  j} \;,
\end{equation}
where we expanded the wavefunction $\Psi_\lambda = \sum\limits_j
\psi_{\lambda  j} \rho_j$ in a fixed basis $\rho_j$ in the dot and
defined  the channel vector
$\phi_{c j} \equiv \left( {\hbar^2 k_c P_c / m } \right)^{1/2}
\int\limits_{\cal  C}dl\;
   \rho^\ast_j(\bbox{r}) \phi_c(\bbox{r})$. We note that the  scalar product in (\ref{scalar-product})
is defined over the dot-lead interface and is different from the
usual scalar product in the Hilbert space of the dot.

\subsection{Peak heights distributions}

\begin{figure}[h!]
\epsfxsize= 7 cm
\centerline{\epsfbox{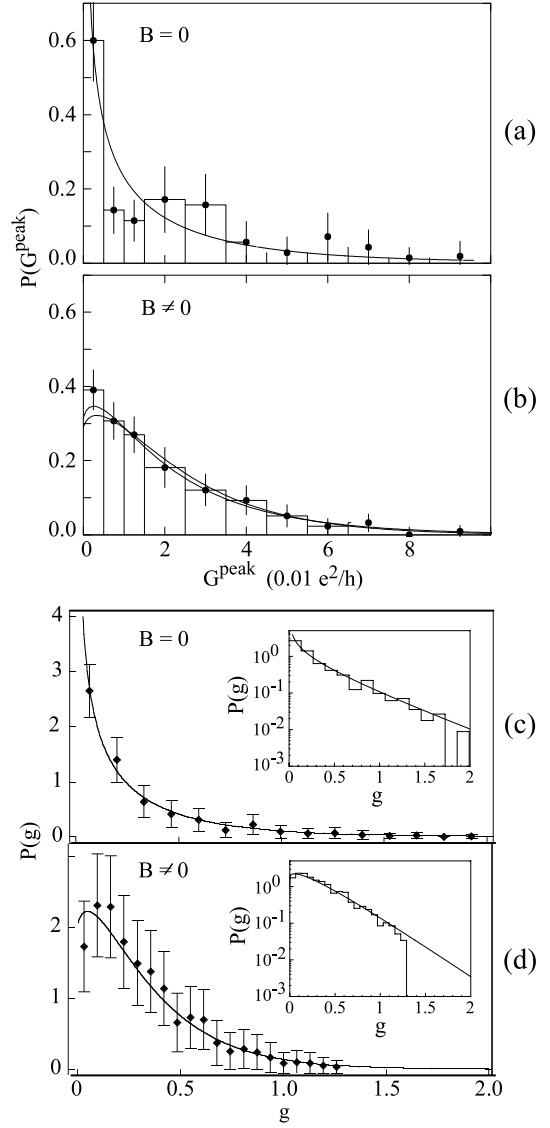}}
\vspace{3 mm}
\caption{Conductance peak-height distributions in Coulomb-blockade quantum dots: (a,b) measured distributions by Chang {\em et al.} \protect\cite{Chang96} for (a) $B=0$ and (b) $B \neq 0$. The effective size of the dot is $0.25\times 0.25$ $\mu$m and $T=75$ mK (corresponding to ${\cal N} \sim 100$ electrons and $T/\Delta \sim 0.15$). The solid lines are the RMT predictions of Jalabert, Stone and Alhassid \protect\cite{Jalabert92}; (c,d) measured distributions of Folk {\em et al.}\protect\cite{Folk96}
 compared with the RMT distributions (solid lines). The dots used have areas of $0.32$ and $0.47$ $\mu$m$^2$ and $T = 70 \pm 20$ mK (${\cal N} \sim 1000$ and $T/\Delta \sim 0.3 -0.5$).  Adapted from Refs. \protect\cite{Chang96} and \protect\cite{Folk96}
}
\label{fig4}
\end{figure}

  The conductance peak distributions for one-channel symmetric leads ($\Lambda=1; \;
\bar{\Gamma}^l  = \bar{\Gamma}^r$) were derived by Jalabert, Stone and Alhassid  using RMT \cite{Jalabert92}. Similar results were obtained in the supersymmetry method \cite{Prigodin93}. Using Eqs. (\ref{peak-height}) and (\ref{scalar-product}), and assuming $\bf \psi_\lambda$ to be an RMT eigenvector, we can calculate the distributions of the dimensionless conductance peak heights $g$.  These distributions  are universal and depend only on the space-time symmetries \cite{Jalabert92,Prigodin93}:
\begin{mathletters}
\begin{eqnarray}\label{Pg-closed}
P_{\rm GOE}(g) & = &  \sqrt{{2 / \pi g}}  e^{-2 g} \label{PgGOE} \\
P_{\rm GUE}(g) & = & 4 g  e^{-2g}  \left[K_0(2g) +K_1(2g) \right]
\;, \label{PgGUE}
\end{eqnarray}
\end{mathletters}
where $K_0$ and $K_1$ are 
modified  Bessel functions. The distributions were measured independently by Chang {\em et al.} \cite{Chang96}  and by Folk {\em et al.} \cite{Folk96} and found to agree with the theoretical predictions for both conserved and broken time-reversal symmetries. Fig. \ref{fig4} shows a comparison between the theoretical and experimental distributions in both experiments. The peak height distributions in closed dots with  several possibly correlated and inequivalent channels in each lead were derived in Refs. \cite{Mucciolo95,Alhassid95'}.

\subsection{Weak localization}

  Another signature of chaos is the weak localization effect. This well-known quantum interference effect was already observed in macroscopic disorder conductors \cite{Bergmann84}.  When time-reversal symmetry is conserved, time-reversed orbits contribute coherently to enhance the return probability. The average conductance is then smaller in the absence of magnetic field. 

\begin{figure}[h!]
\epsfxsize= 6.5 cm
\centerline{\epsffile{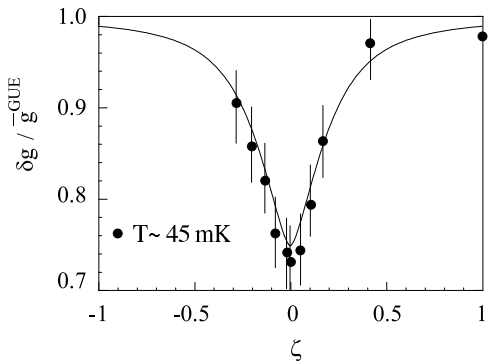}}
\vspace{3 mm}
\caption{ Weak localization in Coulomb-blockade dots. The analytical prediction (\protect\ref{wlocal1}) for the average conductance $\bar g/\bar g^{\rm GUE}$ vs a time-reversal symmetry breaking parameter $\zeta$ (solid line) \protect\cite{Alhassid98''} is compared with the experimental results of Ref. \protect\cite{Folk00} (solid circles). In the experiment $\zeta= B/B_{\rm cr}$  is a scaled magnetic field ($ B_{\rm cr} \approx  6$ mT), and $\bar g^{\rm GUE}$ is measured for $B$ away from zero.
}
\label{fig5}
\end{figure}

  In Coulomb-blockade quantum dots, the average conductance peak height can be calculated as a function of a time-reversal symmetry breaking parameter $\zeta$. In RMT, the statistics of the crossover regime between GOE and GUE can be described by the Mehta-Pandey  ensemble \cite{Mehta83}
\begin{equation}\label{transition-ensemble}
H= S+i\alpha A \;,
\end{equation}
where $S$ and $A$ are, respectively, symmetric and antisymmetric real matrices
and $\alpha$ is a real parameter. The matrices $S$ and $A$ are uncorrelated and chosen from Gaussian ensembles of the same variance. The transition  parameter $\zeta$ is given by a typical symmetry-breaking matrix element
measured  in units of $\Delta$ \cite{French82}:
$\zeta = (\overline{H^2}_{\rm break})^{1/2}/\Delta = {\alpha
\sqrt{N}  / \pi}$ (where $N$ is the dimension of the random matrix). If the time-reversal symmetry is broken by a magnetic field $B$, then $\zeta=B/B_{\rm cr}$, where $B_{cr}$ is the crossover field.

Using the ensemble (\ref{transition-ensemble}) to describe the statistics of the partial amplitude (\ref{scalar-product}), we can find the average conductance peak height in a dot with single-channel symmetric leads \cite{Alhassid98''}
\begin{eqnarray}\label{wlocal1}
\bar{g}(\zeta) = \frac{1}{4} +  & & \int_0^1 d\zeta P_\zeta(t) \times  \nonumber \\ & & \left[
 \left(\frac{t}{1-t^2}\right)^2 \left({2 t^2 \over 1- t^4} \ln t
+\frac{1}{2}  \right)
\right]  \;.  
\end{eqnarray}
Eigenvectors in the crossover ensemble (\ref{transition-ensemble}) have complex components whose distribution in the complex plane is characterized by a ``shape'' parameter $t$.  The function $P_\zeta(t)$ in (\ref{wlocal1}) describes the probability of an eigenvector to have a certain shape $t$ and is given by \cite{Falko94,vanlangen97,Alhassid98}
\begin{eqnarray}\label{shape-dist}
 P_\zeta(t) =  \pi^2 { 1-t^4 \over t^3} \zeta^2 & &
 e^{-{\pi^2 \over 2} \zeta^2 \left(t- 1/t \right)^2} 
 \left\{ \phi_1(\zeta) +   \left[ \frac{1}{4} \left( t + \frac{1}{t}
\right)^2  \right. \right.  \nonumber \\& & \left. \left.   -  {1 \over 2\pi^2 \zeta^2 } \right]  \left[ 1 - \phi_1(\zeta) \right] \right\}  \;,
\end{eqnarray}
where $\phi_1(\zeta)= \int\limits_0^1
 e^{ - 2 \pi^2 \zeta^2 (1 - y^2)} dy$.

  The average dimensionless conductance $\bar g (\zeta)$ has a dip at $\zeta=0$ (see solid line in Fig. \ref{fig5}), describing a weak localization effect.  The theoretical results compare well with recent experimental results vs magnetic field \cite{Folk00} (solid circles in Fig. \ref{fig5}) once the magnetic field is scaled by $B_{cr} \approx 6$ mT. 

  The crossover field can be estimated semiclassically.  Time-reversal symmetry is fully broken for field $B_{cr}$  where the rms of the phase difference between an orbit and its time-reversed partner is $\sim 2\pi$. This phase is proportional to the area enclosed by the electron's trajectory.  In a chaotic dot, area accumulation is diffusive, and the accumulated area's  rms behaves as the squared-root of the elapsed time \cite{Jensen91}. In an open dot the relevant time is the escape time $\tau_{\rm escape}$, but in a closed dot this time is replaced by the Heisenberg time $\tau_H=h/\Delta$.  A more quantitative derivation (for a dot with area ${\cal A}$) gives 
\begin{equation}\label{B-correlation-closed}
{B_{cr} {\cal A} / \Phi_0} = \kappa \left(2\pi{\tau_c / \tau_H}\right)^{1/2}=
\kappa   g_T^{-1/2} = \kappa ({4\pi^2) \cal N}^{-1/4}
\;,
\end{equation}
where $\tau_c$ is the ergodic time (roughly the time to cross the dot), and $g_T=E_T/\Delta$ (with $E_T\equiv \hbar/\tau_c$) is the ballistic Thouless conductance. The factor $\kappa$ is a non-universal geometrical factor.

\subsection{Parametric correlations}\label{sec:parametric}

  Another signature of quantum chaos are the mesoscopic fluctuations of a given conductance peak height as a function of an external parameter, e.g., the shape of the dot or a magnetic field.  These fluctuations  can be described in the framework of Gaussian processes (GP) \cite{Alhassid95} which generalize the random-matrix ensembles to random-matrix processes. A Gaussian process $H(x)$  of a given symmetry class $\beta$ is characterized by its first two moments
\begin{eqnarray} \label{GP-correlations}
   \overline {H_{ij} (x) }  =  0 \;;\;\;\;
   \overline{ H_{ij} (x) H_{kl} (x^\prime) }   =  {a^2 \over {2\beta}}
   f(x - x^\prime) g^{(\beta)}_{ij,kl} \;,
\end{eqnarray}
where the coefficients $g^{(\beta)}_{ij,kl}$ are defined by
\begin{eqnarray}
   g^{(\beta=1)}_{ij,kl} & = & \delta_{ik}\delta_{jl}+\delta_{il}\delta_{jk}
   \;;\;\;\;\;
   g^{(\beta=2)}_{ij,kl} = 2\delta_{il}\delta_{jk}\;. \label{gbeta}
\end{eqnarray}
The Gaussian process (\ref{GP-correlations})
constitutes a Gaussian ensemble for each value of the parameter $x$. 

\begin{figure}[h!]
\epsfxsize= 8. cm
\centerline{\epsffile{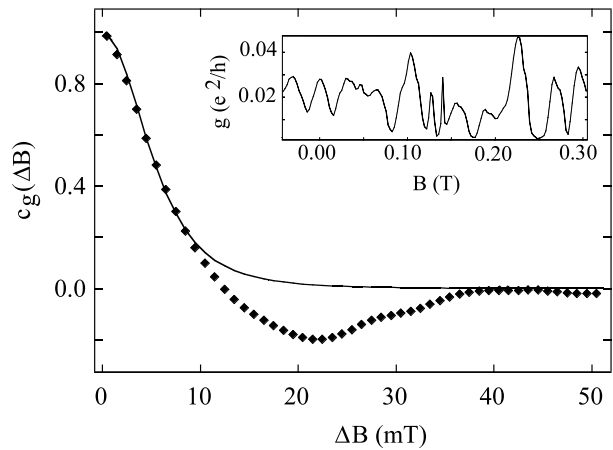}}
\vspace{3 mm}
\caption{ The conductance peak height correlator $c_g$ vs magnetic field $\Delta B$  in closed dots. The experimental results of Ref. \protect\cite{Folk96} (solid diamonds) are compared with the theoretical prediction (\protect\ref{GUP-corr}) of Ref. \protect\cite{Alhassid96} (solid line). The inset shows the measured peak height of a single Coulomb-blockade peak vs magnetic field $B$.  Adapted from Ref. \protect\cite{Folk96}.
}
\label{fig6}
\end{figure}

   A GP is characterized by the short distance behavior of its correlation function $f(x-x') \sim 1-\kappa |x-x'|^\eta$. A differentiable GP is obtained for $\eta=2$ \cite{Attias95}, and corresponds to the usual situation where the Hamiltonian depends analytically on  the parameter. Simons and Altshuler \cite{Simons93} showed that parametric correlations in disordered or chaotic systems become universal once the parameter $x$ is scaled by the rms of the level velocity \cite{Simons93}
\begin{eqnarray}\label{scaling}
   x\rightarrow\bar{x}\equiv 
\left[\overline{\left({{\partial\epsilon_i}/{\partial
x}}\right)^2}\right]^{1/2}  x \;.
\end{eqnarray}
Here $\epsilon_i$ is the $i$th energy level measured in units of the mean-level spacing  $\Delta$. 
 The conductance peak correlator is defined by $c_g( x - x') = \overline {\delta G(x)\delta G(x^\prime)}/
[\sigma(G(x)) \sigma(G(x'))]$, where  $\delta G(x) = G(x) - \bar G(x)$ and  $\sigma^2(G(x))=
\overline{(\delta  G(x))^2}$ (here $G(x)$ is the conductance peak height at a value $x$ of the parameter). The universal correlator $c_g$
 was calculated in Ref. \cite{Alhassid96} using the simple GP \cite{Wilkinson92,Alhassid95} $H(x) = H_1 \cos x + H_2 \sin x$, where $H_1, H_2$ are uncorrelated matrices that belong to the appropriate Gaussian ensemble.  The GUE correlator can be well fitted to the square of a Lorentzian. If the parameter is a magnetic field $B$ then 
\begin{equation}\label{GUP-corr}
c_g(\Delta B) \approx \left[ {1 \over 1+ (\Delta B/B_c)^2} \right]^{2} \;,
\end{equation}
where $B_c$ is the correlation field. Fig. \ref{fig6} shows the measured conductance peak correlator (solid diamonds) in comparison with the theoretical prediction  (\ref{GUP-corr}) (solid line), where the parameter $B_c$ is fitted. RMT describes the correct shape of the correlator for $B \alt 10 $ mT. A single-particle semiclassical estimate of the correlation field $B_c$ gives an expression similar to (\ref{B-correlation-closed}). For a stadium in a uniform magnetic field, the correlation flux is $\Phi_c \approx 0.3\; \Phi_0$ \cite{Bohigas95,Alhassid98}, which is below the experimental value of $\approx 0.8\;\Phi_0$. This indicates that the single-particle picture is inadequate for estimating the correlation field.  Numerical simulations in small disordered dots with Coulomb interactions find that the correlation field increases with the interaction strength \cite{Berkovits98'}.

\subsection{Peak-to-peak correlations}

The conductance peak at finite temperature ($T \sim \Delta$) can be calculated in the master-equations approach \cite{Beenakker91}
\begin{eqnarray}\label{finite-T-G}
\begin{array}{ll}
 G (T,\tilde{E}_F) = \frac{e^2}{h}\, \frac{\pi \bar{\Gamma}}{4 kT} g\;,
\;\;\;  &
{\rm where} \;\;
 g = \sum_\lambda w_\lambda(T,\tilde{E}_F) g_\lambda \;
\end{array}
\end{eqnarray}
is the dimensionless conductance expressed as a thermal average over the level  conductances
  $g_\lambda =  2 \bar{\Gamma}^{-1}  \Gamma_\lambda^l \Gamma_\lambda^r
   /( \Gamma_\lambda^l + \Gamma_\lambda^r)$.
For $T,\Delta \ll e^2/C$, the thermal weights for the ${\cal N}$-th conductance peak are given by
\begin{eqnarray}\label{thermal-weight'}
w_\lambda =  4 P_{\cal N} \langle n_\lambda \rangle_{_{\cal N}}
\left[
1 - f\left(E_\lambda  - \tilde{E}_F \right)
\right]  \;,
\end{eqnarray}
where $P_{\cal N}$ is the probability that the dot has ${\cal N}$ electrons,
and  $\langle n_\lambda \rangle_{_{\cal N}} $ is the canonical
occupation  of a level $\lambda$.

\begin{figure}[h!]
\epsfxsize= 8. cm
\centerline{\epsffile{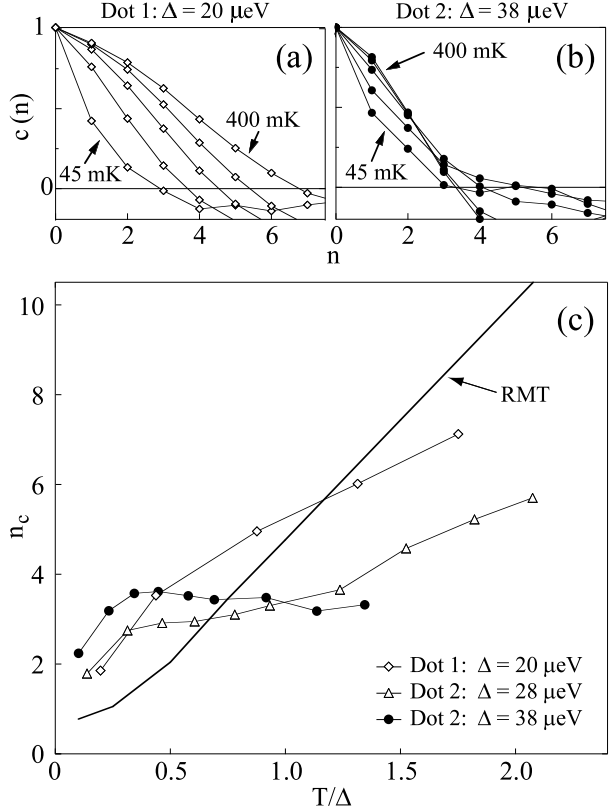}}
\vspace{3 mm}
\caption{Finite-temperature peak-to-peak correlations (experiment). The peak-to-peak correlator $c(n)$ is shown at different temperatures for (a) a larger dot with $\Delta =20$ $\mu$eV and 
(b) a smaller dot with $\Delta =38$ $\mu$eV. (c) The number of correlated peaks $n_c$ (defined as the FWHM of $c(n)$) vs $T/\Delta$ for three dots. The solid line is the RMT result for an unchanged spectrum. From Ref. \protect\cite{Patel98}.
}
\label{fig7}
\end{figure}

The finite-temperature peak height statistics were calculated in Ref. \cite{Alhassid98'} assuming  the level conductance and energy levels satisfy RMT. Of particular interest is the peak-to-peak  correlator
\begin{equation}\label{peak-to-peak}
c(n) =  { \overline{\delta G _{\cal N} \delta G_{{\cal N}+n}} /
 \overline{(\delta G_{\cal N})^2} } \;,
\end{equation}
where $\delta G_{\cal N}= G_{\cal N}- \bar G_{\cal N}$ is the fluctuation of
the  ${\cal N}$-th conductance peak around its average. The number of correlated peaks $n_c$ is the FWHM of (\ref{peak-to-peak}). The calculated temperature dependence of $n_c$ is shown in  Fig. \ref{fig7}(c) (solid line). 

  The experimental results \cite{Patel98} shown in Fig. \ref{fig7} demonstrate that $n_c$ saturates with temperature, especially for the small dots. This behavior is contrary to the linear dependence predicted by RMT. An explanation of this saturation effect is discussed in Sec. \ref{sec:scrambling}.

\section{Interaction effects}\label{sec:interactions}

 While the CI-plus-RMT model can explain some of the observed statistical properties of closed dots,  several experimental observations suggest that electron-electron interactions beyond the charging energy are important:
(i) The measured peak-spacing distribution \cite{Sivan96,Simmel97,Patel98'} does not have the Wigner-Dyson shape expected in the CI model (see Eq. (\ref{peak-spacing})). This signature will be discussed in detail in Sec. \ref{sec:peak-spacings}.
(ii) The measured correlation flux is larger than the semiclassical estimate (see Sec. \ref{sec:parametric}). (iii) Correlations between the addition and excitation spectra are seen only for a small number of added electrons \cite{Stewart97}. (iv) The peak-to-peak correlations saturate with increasing temperature, contrary to the results of the CI-plus-RMT model.   This effect is explained in Section \ref{sec:scrambling}.

\subsection{Spectral scrambling}\label{sec:scrambling}

 The best way to include interaction effects while retaining a single-particle framework is the  mean-field approach, e.g., the Hartree-Fock approximation. In this approach, the self-consistent single-particle energy levels change (``scramble'') when an electron is added to the dot.  This scrambling can be modeled in terms of a discrete parametric random matrix $H(x_{\cal N})$, where $x_{\cal N}$ describes the parameter of the dot with ${\cal N}$ electrons.  We describe $H(x_{\cal N})$  as a discrete Gaussian process, and embed it in a continuous process $H(x)$.  We assume that the scaled parametric change $\delta \bar x = \bar x_{{\cal N}+1} - \bar x_{\cal N}$ upon the addition of one electron is independent of  ${\cal N}$. 

\begin{figure}[h!]
\epsfxsize= 7. cm
\centerline{\epsffile{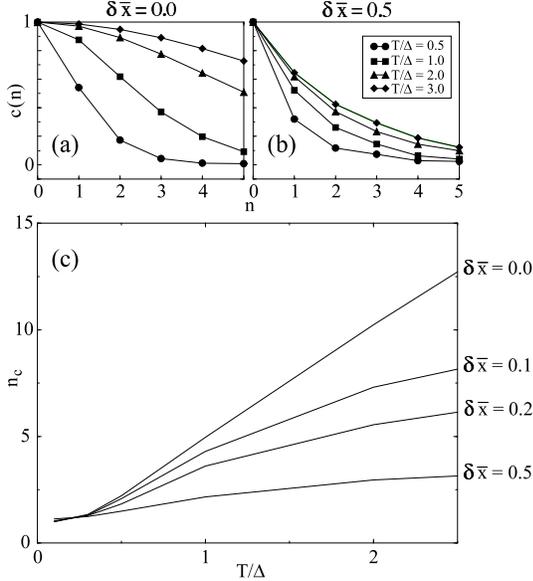}}
\vspace{3 mm}
\caption{ Finite-temperature peak-to-peak correlations and spectral scrambling (theory). The calculated correlators $c(n)$ are shown at several temperatures for (a) a dot with an unchanged spectrum ($\delta \bar x=0$), and (b) a dot where the spectrum scrambles ($\delta \bar x=0.5$). (c) The number of correlated peaks $n_c$ vs $T/\Delta$ for dots with different scrambling parameters $\delta \bar x$. Notice the earlier saturation of $n_c$ as the spectrum scrambles faster (i.e., when $\delta\bar x$ is larger). From Ref. \protect\cite{Alhassid99}.
}
\label{fig8}
\end{figure}

  The parameter $\delta \bar x$ determines the degree of spectral scrambling. Fig. \ref{fig8}(a)(b) shows the correlators $c(n)$ at various temperatures for both an unchanged spectrum ($\delta \bar x=0$) and for a spectrum that scrambles ($\delta \bar x=0.5$). Fig. 8(c) shows that when $\delta \bar x$ is larger (i.e., the spectrum scrambles faster), the number of correlated peaks saturates at smaller values of $T/\Delta$ \cite{Alhassid99}.  

Spectral scrambling is an interaction effect \cite{Patel98}. A microscopic estimate of $\delta\bar x$ can be done in Koopmans' limit \cite{Koopmans}, i.e., assuming the single-particle wave functions do not change with the addition of electrons (and only the spectrum scrambles). In this limit the change $\delta E_i$ of an energy level upon the addition of one electron is given by a diagonal matrix element   $\delta E_i \approx v_{i, {\cal N}+1}$, where $v_{\alpha \gamma} \equiv v_{\alpha \gamma; \alpha \gamma}$. Using an RPA screened interaction \cite{Blanter97} or a short-range dressed interaction \cite{Altshuler97}  $v \sim \lambda \Delta {\cal A} \delta(\bf r - \bf r')$, we have
\begin{equation}\label{fluct1}
\sigma^2(\delta E_i) \approx \sigma^2(v_{i, {\cal N}+1}) \sim {\lambda^2 \over \beta^2} { \Delta^2 \over g_T^2} \;,
\end{equation}
where $g_T$ is the Thouless conductance. In the finite dot, there could be an additional contribution due to excess negative charge on the boundaries \cite{Blanter97}, leading to 
\begin{equation}\label{fluct2}
\sigma^2(\delta E_i) \sim {\lambda^2 \over \beta} {\Delta^2 \over g_T} \;.
\end{equation}
  In the parametric approach $\sigma^2(\delta E_i) \approx (\delta \bar x)^2$, and by comparing with (\ref{fluct1}) or (\ref{fluct2}) we can find $\delta \bar x$.  The number of added electrons $m$ required for complete scrambling of the spectrum is then determined from $m_c \delta \bar x \sim 1$. We obtain \cite{Alhassid99,Alhassid00c}
\begin{mathletters}\label{Delta-x-g}
\begin{eqnarray}
& m_c \sim {\beta g_T \over \lambda} \sim {\beta  {\cal N}^{1/2}\over \lambda} \label{Delta-x-g1}\\
{\rm or}\;\;\; & m_c \sim {(\beta g_T)^{1/2}\over \lambda} \sim {\beta ^{1/2} {\cal N}^{1/4}\over \lambda}\;,\label{Delta-x-g2} 
\end{eqnarray}
\end{mathletters}
where (\ref{Delta-x-g2}) holds in the presence of surface charge.

\subsection{Peak spacing statistics}
\label{sec:peak-spacings}

\begin{figure}
\epsfxsize= 8. cm
\centerline{\epsffile{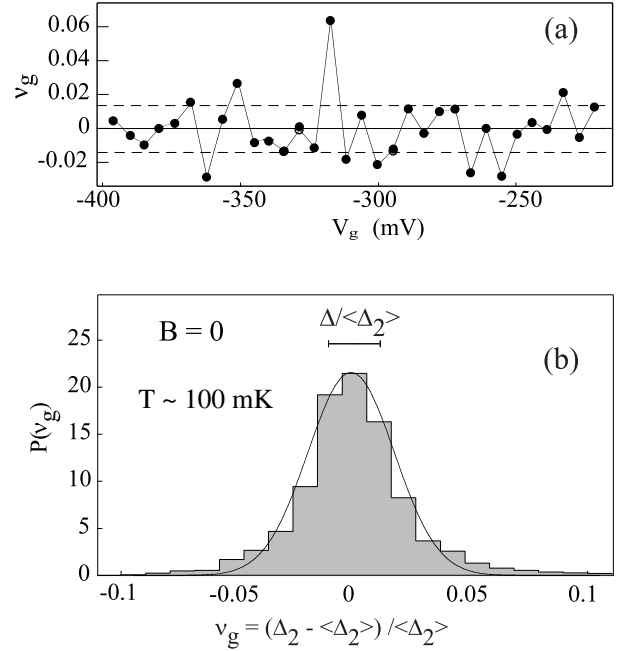}}
\vspace{3 mm}
\caption{Measured peak-spacing statistics in quantum dots: (a) a sequence of  peak-spacing fluctuations $\nu_g \equiv (\Delta V_g -\overline{\Delta  V_g})/ \overline{\Delta V_g}$ vs.~$V_g$ for
$B=30  \;$mT. The dashed lines show the standard deviation of RMT. (b) The measured peak-spacing distribution $P(\nu_g)$ (shaded histograms) for $B=0$ and $T/\Delta \sim 0.8$ is compared with a Gaussian fit (solid line). From Ref. \protect\cite{Patel98'}.
}
\label{fig9}
\end{figure}

One of the main signatures of interactions is seen in the peak spacing distribution. The spacing between peaks in an interacting dot is given by
\begin{equation}\label{spacing-int}
\Delta_2 = {\cal E}_{\rm g.s.}({\cal N}+1) + {\cal E}_{\rm g.s.}({\cal N}- 1)-2 {\cal E}_{\rm g.s.}({\cal N})  \;,
\end{equation}
where ${\cal E}_{\rm g.s.}({\cal N})$ is the ground state energy of the dot with ${\cal N}$ electrons. In the CI model, $\Delta_2$ reduces to (\ref{peak-spacing}), and a shifted Wigner-Dyson  distribution is expected for $\Delta_2$.  Experimentally, the distribution is Gaussian-like (for semiconductor dots with $r_s \sim 1-2$) and has a larger width than the Wigner-Dyson  distribution \cite{Sivan96,Simmel97,Patel98'}.  An example of the measured peak spacing statistics is shown in Fig. \ref{fig9}.

  An estimate of the fluctuations of $\Delta_2$ can be obtained in Koopmans' limit \cite{Blanter97} where the addition energy is given by $E^{({\cal N})}_{\cal N}$ ($E^{({\cal N})}_i$ is the $i$-th single-particle level of a dot with ${\cal N}$ electrons). The peak spacing can then be written as
\begin{equation}\label{spacing-K}
\Delta_2 \approx E^{({\cal N}+1)}_{{\cal N}+1} - E^{({\cal N})}_{\cal N} = (E^{({\cal N}+1)}_{{\cal N}+1} -E^{({\cal N}+1)}_{\cal N}) + \Delta E_{\cal N} \;.
\end{equation}
The first term on the r.h.s. of (\ref{spacing-K}) is the usual level spacing in a dot with a fixed number of electrons (${\cal N}+1$) and has an rms of order $\Delta$. The second term $\Delta E_{\cal N}$ represents the change of a given energy level when an electron is added and its standard deviation can be estimated from (\ref{fluct1}) or (\ref{fluct2}). 

 Gaussian-like peak spacing distributions were explained in exact numerical diagonalization of a small Anderson model with Coulomb interactions \cite{Sivan96,Berkovits98}. Their width increases with the gas constant $r_s$. Scrambling of the single-particle spectrum (see Section \ref{sec:scrambling}) can also lead to Gaussian distributions if $\delta \bar x$ is sufficiently large \cite{Vallejos98}.

Nearly-Gaussian distributions were observed in Hartree-Fock  calculations at larger values of $r_s$ \cite{Levit99,Walker99,Cohen99}. The distribution of a diagonal interaction matrix element is found to be approximately Gaussian. Using (\ref{spacing-K}) (for small values of $r_s$), we can describe the peak spacing distribution as a convolution of the Wigner-Dyson distribution with a Gaussian distribution. 

  The numerical investigations of small disordered dots with interactions demonstrate the need to go beyond the simple CI model. An interesting question is whether it is possible to describe the RMT-like behavior of the peak height statistics and the Gaussian-like distribution of the peak spacing distribution within a single random-matrix model. 

\subsection{Random interaction matrix model}

RMT is not restricted to single-particle systems. It was successfully applied to strongly interacting systems, e.g., the compound nucleus at finite excitations.  However, in the linear conductance regime of quantum dots we are interested in the statistical properties of the ground state of the system as  the number of electrons changes.  Since RMT does not make explicit reference to interactions or to particle number, it is necessary to use a model that contains interactions explicitly. A two-body random-interaction model was introduced in nuclear physics in the early 1970s \cite{French70,Bohigas71}. It was used, together with a random single-particle spectrum, to study thermalization \cite{Flambaum96} and the onset of chaos in interacting many-body systems \cite{Jacquod97}. Since its one-body part has Poissonian statistics it is not suitable for studying chaotic dots. A random interaction matrix model (RIMM) for dots with chaotic single-particle dynamics was recently introduced to study the interplay between one-body chaos and interactions \cite{Alhassid00}.  

The RIMM describes an ensemble of interacting Hamiltonians  in a fixed basis of $m$ single-particle states
$|i\rangle  = a_i^\dagger |0\rangle$
\begin{eqnarray}\label{RIMM}
H = \sum\limits_{ij} h_{ij} a^\dagger_i a_j +{1\over 4} \sum_{ijkl}
 u^A_{ijkl}a^\dagger_i a^\dagger_j a_l a_k
\;.
\end{eqnarray}
The one-body elements $h_{ij}$ are chosen from the appropriate
 Gaussian random-matrix ensemble, i.e., GOE (GUE) for conserved (broken) time-reversal symmetry of the one-body dynamics.  The anti-symmetrized two-body matrix elements $u^A_{ij;kl}  \equiv u_{ij;kl} - u_{ij;lk}$
form a GOE in the two-particle space (the two-body interaction is assumed to conserve time-reversal symmetry irrespective of the symmetry of the one-body Hamiltonian).
 The variance of the diagonal (off-diagonal) interaction matrix elements is $U^2$ ($U^2/2$).    The two-body interaction can include a non-vanishing average part $\bar u$ that is invariant under orthogonal transformations of the single-particle basis.  The only such invariant for spinless electrons is the charging energy $e^2 {\cal N}^2/2C$, which is a constant and thus does not affect the statistical fluctuations. If the spin degrees of freedom are included in the RIMM then another possible invariant is the exchange interaction $-\xi \bbox S^2/2$ \cite{Kurland00}. It is important to note that in a physical model of the dot, the two-body interaction in a given basis is fixed. The introduction of a fluctuating two-body interaction is done in the spirit of RMT to describe generic effects that do not depend on the specific interaction. 

\begin{figure}
\epsfxsize= 7. cm
\centerline{\epsffile{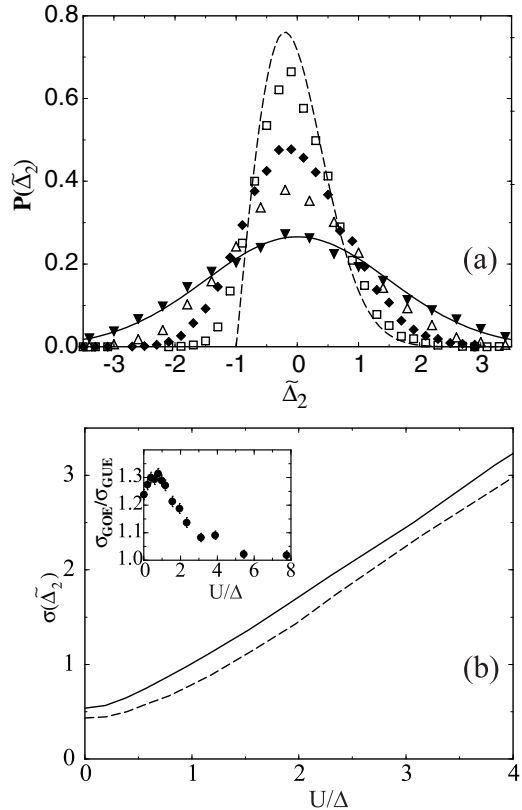}}
\vspace{3 mm}
\caption{Peak spacing statistics in the RIMM: (a) $B=0$ peak-spacing distribution for several values of $U/\Delta$: 0.35 (open squares), 0.7 (solid diamonds), 1.1 (open triangles) and 1.8 (solid triangles). The dashed line is the Wigner-Dyson distribution and the solid line is a Gaussian fit to the $U/\Delta=1.8$ distribution. (b) The standard deviation $\sigma(\Delta_2)$ vs $U/\Delta$ for GOE (solid line) and GUE (dashed line) one-body statistics. The inset is the ratio $\sigma_{\rm GOE}(\Delta_2)/\sigma_{\rm GUE}(\Delta_2)$ vs $U/\Delta$. All the results are shown for $m=12$ and ${\cal N}=4$.
Adapted from Ref. \protect\cite{Alhassid00}.
}
\label{fig10}
\end{figure}

  The RIMM was used to calculate the peak spacing distribution from Eq. (\ref{spacing-int}), where the ground state energies of the Hamiltonian (\ref{RIMM}) are calculated for ${\cal N}-1$, ${\cal N}$ and ${\cal N}+1$ electrons. The peak spacing distribution describes a crossover from a Wigner-Dyson distribution at $U=0$ to a Gaussian distribution as $U/\Delta$ increases; see Fig. \ref{fig10}(a) \cite{Alhassid00}.  The width of the distributions increases vs $U/\Delta$ (see Fig. \ref{fig10}(b)). The distributions are well described by a  convolution of a Wigner-Dyson distribution with a Gaussian distribution. For small $U/\Delta$, the width of this Gaussian is just the standard deviation of a diagonal interaction matrix element. 

  The inset of Fig. \ref{fig10}(b) shows the calculated ratio $\sigma_{\rm GOE}(\Delta_2)/\sigma_{\rm GUE}(\Delta_2)$ vs $U/\Delta$. The experimental values  $\sim 1.1-1.3$ \cite{Patel98'} are consistent with theory. 

\begin{figure}
\epsfxsize= 8 cm
\centerline{\epsffile{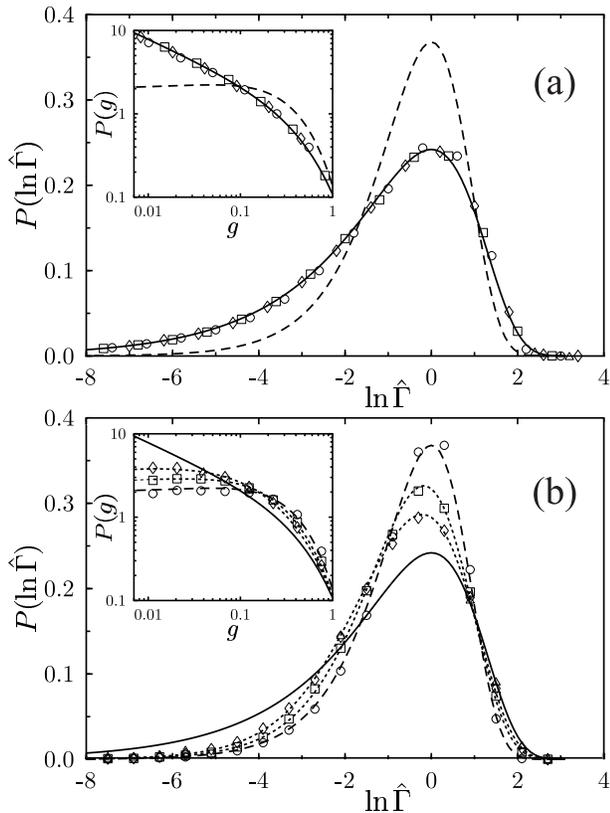}}
\vspace{3 mm}
\caption{Width and conductance peak height statistics in the RIMM: (a) Width distributions $P(\ln \hat \Gamma)$ vs.~$\ln \hat \Gamma$ are shown for GOE one-body statistics and $U/\Delta=0$
(circles), $2.4$ (squares), and $4$ (diamonds). The solid and dashed lines are the GOE and GUE Porter-Thomas distributions, respectively.
(b) Same as in (a) but for GUE one-body statistics.   The short-dashed lines are analytic width distributions derived from the crossover ensemble (\protect\ref{transition-ensemble}). All the results are shown for $m=12$ and ${\cal N}=4$.
 The  insets in (a) and (b) are the corresponding peak-height distributions $P(g)$ in a log-log scale. Adapted from Ref. \protect\cite{Alhassid00b}.
}
\label{fig11}
\end{figure}

  To calculate the conductance peak height we use Eq. (\ref{peak-height}) but now the partial width of the ground state of
the ${\cal N}$-electron dot to decay into the ground state of the dot with ${\cal N}-1$ electrons plus an electron in the respective lead is given by:
\begin{equation}\label{width}
\Gamma_{\cal N} \propto \left|\langle \Phi_{\rm g.s.}({\cal N}) |
 \;\psi^\dagger({\bbox r}) |\; \Phi_{\rm g.s.}({\cal N}-1) \rangle \right|^2 \;,
\end{equation}
where $\psi^\dagger({\bbox r})$ is the creation operator of an
electron at the point contact $\bbox r$, and $\Phi_{\rm
g.s.}({\cal N})$ is the ground state wavefunction of the ${\cal N}$-electron dot.  For a GOE one-body statistics we find that the partial-width distribution is a Porter-Thomas distribution irrespective of $U/\Delta$ (see Fig. \ref{fig11}(a)). A similar insensitivity to the interaction strength was found in numerical simulations of an Anderson model with Coulomb interactions \cite{Berkovits98'}. However, for a GUE single-particle statistics (Fig. \ref{fig11}(b)) we find a crossover from the GUE  Porter-Thomas distribution (at U=0) to the GOE Porter-Thomas distribution (at large $U/\Delta$).  Equivalently, the GOE $\to$ GUE transition due to an external magnetic field is incomplete because of the competing GOE symmetry of the two-body interaction.  The crossover width distributions are well described by those of the Mehta-Pandey ensemble (\ref{transition-ensemble}) with a transition parameter $\zeta$ that is a monotonically decreasing function of $U/\Delta$ \cite{Alhassid00b}. 

  We note that the curves of the width $\sigma(\Delta_2)$ and $\zeta$ vs $U/\Delta$ depend on both $m$ (number of single-particle states) and ${\cal N}$ (number of electrons), but they become universal once $U/\Delta$ is scaled by a constant (that depends on $m$ and ${\cal N}$). For $U/\Delta \sim 0.7 - 1.5$, the peak spacing distribution is already Gaussian-like, while the peak height statistics (in the presence of a time-reversal symmetry breaking field) is still close to the GUE prediction.  In the RIMM, $U/\Delta$ is a free parameter and physical values can be determined by comparing its results to specific models. For a small ($4\times 5$) Anderson model with Coulomb interactions of strength $U_c$, we find that the range $U/\Delta \sim 0.7 - 1.5$ corresponds to $U_c \sim 2 - 5$ \cite{Alhassid00,Alhassid00b}. 

\section{Conclusions}

The mesoscopic fluctuations in closed dots are affected by both one-body chaos and electron-electron interactions.  Signatures of chaos include the RMT-like conductance peak height distributions, the weak localization effect, and the line shape of the parametric peak-height correlator.  Interaction effects include the Gaussian-like shape of the peak-spacing distribution, a correlation magnetic field that is larger than its single-particle estimate, and the saturation of peak-to-peak correlations with temperature. Some of these interaction effects can be described by a random interaction matrix model.

\section{Acknowledgments}
 
 I thank H. Attias, Y.Gefen, M. G\"{o}k\c{c}eda\u{g}, J.N. Hormuzdiar,  Ph. Jacquod, R. Jalabert, S. Malhotra, C.M. Marcus, S.  Patel, A.D. Stone, N. Whelan and A. Wobst for collaborations on various parts of the work presented above.  This research was supported in part by the Department of Energy grant No. DE-FG-0291-ER-40608.

\end{document}